\documentclass{ieeeaccess}
\usepackage{cite}

\usepackage{graphicx}
\usepackage{physics}
\usepackage{amsmath,amssymb,amsfonts,mathtools}
\usepackage{fancyhdr}
\usepackage{comment}
\usepackage[absolute,overlay]{textpos}

\newcommand\identity{1\kern-0.25em\text{l}}

\begin{document}
\doi{} 


\title{Pre-amplification for Preserving Quantum Resources Through an Open-Air, Room-Temperature Microwave Channel}

\author{\uppercase{Jonathan N. Blakely}\authorrefmark{1}, \IEEEmembership{Senior Member, IEEE},
\uppercase{Drew S. Addison}\authorrefmark{2}, and \uppercase{Mark L. Adams}\authorrefmark{2},
\IEEEmembership{Senior Member, IEEE}}
\address[1]{Department of Electrical and Computer Engineering, University of Alabama in Huntsville, Huntsville, Alabama 35899, USA}
\address[2]{Department of Electrical and Computer Engineering, Auburn University, Auburn, Alabama 36849 USA}

\markboth
{Blakely \headeretal: Pre-amplification for Preserving Quantum Resources}
{Blakely \headeretal: Pre-amplification for Preserving Quantum Resources}

\corresp{Corresponding author: Jonathan N. Blakely (email: jonathan.blakely@uah.edu).}

\begin{abstract}
We model a scenario where the pre-amplified signal mode of a two-mode
squeezed vacuum (TMSV) state generated by a Josephson parametric converter (JPC) is transmitted through a room-temperature, open-air channel, while the idler is retained at low temperature. Before transmission, the state contains three quantum resources: (1) entanglement, (2) non-classicality, and (3) sub-vacuum conditional prediction of the signal quadrature. To preserve these resources through noise and loss, we consider two forms of quantum-limited pre-amplification of the signal mode: phase-preserving amplification (PPA) and phase-sensitive amplification (PSA). We identify three thresholds for a fixed amount of loss as noise increases from zero marking the loss of
(1) entanglement, (2) non-classicality, and (3) sub-vacuum conditional prediction of the signal quadrature, respectively. We find first that entanglement
is fatally sensitive to the noise and loss in our channel model and neither form of pre-amplification protects it. On the other hand, phase-sensitive pre-amplification is shown to significantly raise the thresholds at which non-classicality and sub-vacuum conditional prediction are destroyed. For a 3-meter link between room-temperature horn antennas at 5 GHz, the noise thresholds for these resources can be orders of magnitude above the thermal background. Further analysis shows that these conclusions are robust to limited phase misalignment and pre-amplifier-added noise, with the latter having the stronger effect. These surviving non-classical resources may support quantum-enhanced sensing, illumination, and metrology protocols even in regimes where entanglement itself is destroyed. Our results suggest that protocols that exploit these quantum phenomena are the most likely to succeed in microwave applications.\end{abstract}


\titlepgskip=-15pt

\maketitle

\section{Introduction}
Microwave implementation of many quantum protocols is impeded by the large loss and background noise characteristic of room temperature, open-air transmission \cite{weedbrook2012continuous, xiang2017intracity,magnard2020microwave,barzanjeh2020microwave,qasymeh2022high,fesquet2023perspectives, qiu2025thermal,yam2025cryogenic}. Entanglement available in a two-mode squeezed vacuum state generated at millikelvin temperatures is quickly obliterated outside the protective environment of a dilution refrigerator \cite{xiang2017intracity,magnard2020microwave, qasymeh2022high,gonzalez2022open,wang2026thermal}. But is that the end of the story? Previously, it has been shown that noiseless or phase-sensitive pre-amplification can partially mitigate downstream loss and preserve measurable non-classical correlations and squeezing in optical systems \cite{leonhardt1994high,bencheikh1995improvement,bencheikh1997cascaded,lam1997noiseless,fossier2009improvement,ulanov2015undoing,li2017improved}. Here we demonstrate the potential for protecting quantum resources in an open-air, microwave channel by pre-amplification before transmission. 

\Figure[t!](topskip=0pt, botskip=0pt, midskip=0pt)[width=0.90\textwidth]{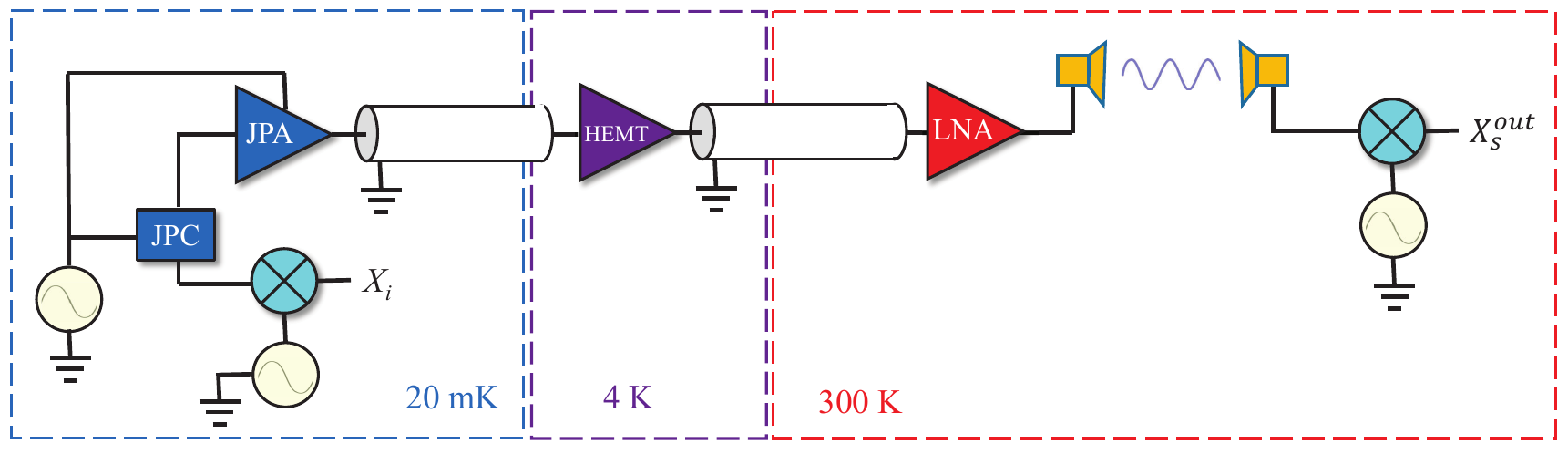}
{\parbox[t]{0.85\textwidth}{Schematic representation of a quantum microwave source, an open-air channel, and a receiver. The TMSV state is generated at low temperature by a Josephson parametric converter (JPC) with signal and idler output modes on separate transmission lines. The signal mode is pre-amplified by a Josephson parametric amplifier (JPA) configured either as a PPA or PSA. The pump oscillator and local oscillators are all derived from an original signal source. A transmission line carries the amplified signal mode out of the low-temperature environment, through an intermediate stage with a conventional HEMT amplifier, and finally to a room-temperature stage with an LNA feeding the antenna for transmission, propagation, reception, and homodyne measurement.}\label{fig:Microwave_Channel_Schematic}}

Specifically, we analyze a scenario where the pre-amplified signal mode of a two-mode squeezed vacuum (TMSV) state generated by a Josephson parametric converter (JPC) is transmitted through a room-temperature, open-air channel, while the idler is retained at low temperature as shown schematically in Fig. \ref{fig:Microwave_Channel_Schematic}. We find that, although entanglement is fatally sensitive to noise and loss, pre-amplification can preserve other resources, namely, non-classicality and sub-vacuum conditional prediction \cite{heidmann1987observation, mallet2011quantum}. Here, we use the term \textit{quantum resource} in its broader operational sense to denote any property of quantum origin that can be exploited to enhance an informational or sensing task, rather than in the stricter axiomatic sense of quantum resource theory \cite{chitambar2019quantum}. These surviving non-classical resources can potentially support quantum-enhanced sensing, illumination, and metrology protocols even in regimes where entanglement itself is destroyed \cite{barzanjeh2015microwave,zhang2015entanglement,zhuang2018distributed,lecocq2015quantum}. Our results suggest that protocols that exploit these quantum phenomena are the most likely to succeed in microwave applications.

The TMSV state, before amplification or transmission, contains all three quantum resources. In the number state representation, the TMSV is
\begin{align}
    |\psi\rangle=\frac{1}{\cosh r}
\sum_{n=0}^{\infty}
(\tanh r)^n \; |n\rangle_i \, |n\rangle_s,
\end{align}
where $r$ is the squeezing parameter, and the subscripts $i$ and $s$ denote the idler and signal modes, respectively. Entanglement can be seen directly in this non-product state and is manifested in the perfect correlation in simultaneous photon number measurements of the two modes. Non-classicality follows directly from entanglement. 

In the two-mode, quadrature state representation, the TMSV takes the form
\begin{align}
    \psi&(x_i,x_s)
=\nonumber \\
&\frac{1}{\sqrt{\pi}}
\exp\!\left[
-\frac{1}{4}\left(
e^{2r}(x_i-x_s)^2
+
e^{-2r}(x_i+x_s)^2
\right)
\right],
\end{align}
where $x_i$ and $x_s$ are measurement outcomes for the first quadrature operators $X_i$ and $X_s$, respectively. Define the minimum mean square error of an estimate of one quadrature of the signal mode given a measurement of the same quadrature of the idler as
\begin{align}
\operatorname{MMSE}(X_s|X_i)
\equiv
\min_{\hat{X}_s(\cdot ))}\mathbb E\!\left[(X_s-\hat{X}_s(X_i))^2\right].
\end{align}
Then, because the TMSV quadrature statistics are jointly Gaussian,
$\hat{X}_s^\text{opt}(X_i)=g_\text{opt}X_i$,
where
\begin{align}
g_\text{opt}
=
\frac{\operatorname{Cov}(X_s,X_i)}
{\operatorname{Var}(X_i)}
=
\tanh 2r,
\label{opt_gain}
\end{align}
which gives
\begin{align}
    \operatorname{MMSE}(X_s|X_i)=
\frac{1}{2\cosh{2r}}<1/2,
\label{TMSV_cond_pre}
\end{align}
meaning sub-vacuum prediction of the signal quadrature conditioned on the idler is possible for any non-zero squeezing, $r>0$. The vacuum value \(1/2\) therefore provides a natural benchmark for conditional prediction. In contrast, for any Glauber-Sudarshan $P$-classical state,
 \begin{align}
    \operatorname{MMSE}(X_s|X_i) \ge 1/2,
\end{align}
(see App. \ref{classical_state_MMSE_bound}).

To preserve these resources through noise and loss, we consider two forms of quantum-limited pre-amplification to prepare the signal mode for transmission: phase-preserving amplification (PPA) and phase-sensitive amplification (PSA) \cite{bergeal2010phase}. Phase-preserving amplification of a single mode with quadrature operators $X$ and $P$ enacts the transformation
\begin{align}
X\to \sqrt{G}X+X_\text{add}, \;\;\; P\to \sqrt{G}P+P_\text{add},
\label{PPA_transform}
\end{align}
where $G$ is the power gain, and $X_\text{add}$ and $P_\text{add}$ are quadratures of an added-noise mode.  If the amplifier is quantum-limited, i.e., the added noise is the minimum required by quantum theory, then the noise mode is in its vacuum state and 
\begin{align}
    \mathrm{Var}(X_\text{add})=\mathrm{Var}(P_\text{add})=\frac{G-1}{2}.
    \label{PPA_noise_var}
\end{align} Thus, PPA boosts both quadratures symmetrically and adds a non-zero amount of noise. Figure \ref{fig:Amplifiers}(a) shows the effect in phase space of PPA on an input vacuum state. In contrast, quantum-limited PSA of a single mode with quadrature operators $X$ and $P$ enacts the transformation
\begin{align}
X\to aX,\;\;\;  P\to a^{-1}P,
\label{PSA_transform}
\end{align}
where $a$ is the PSA gain parameter. Under this transformation, one quadrature is boosted while the other is suppressed. Importantly, no added noise is introduced. Figure \ref{fig:Amplifiers}(b) shows the effect in phase space of PSA on an input vacuum state.

\Figure[b!](topskip=0pt, botskip=0pt, midskip=0pt)[width=0.95\columnwidth]{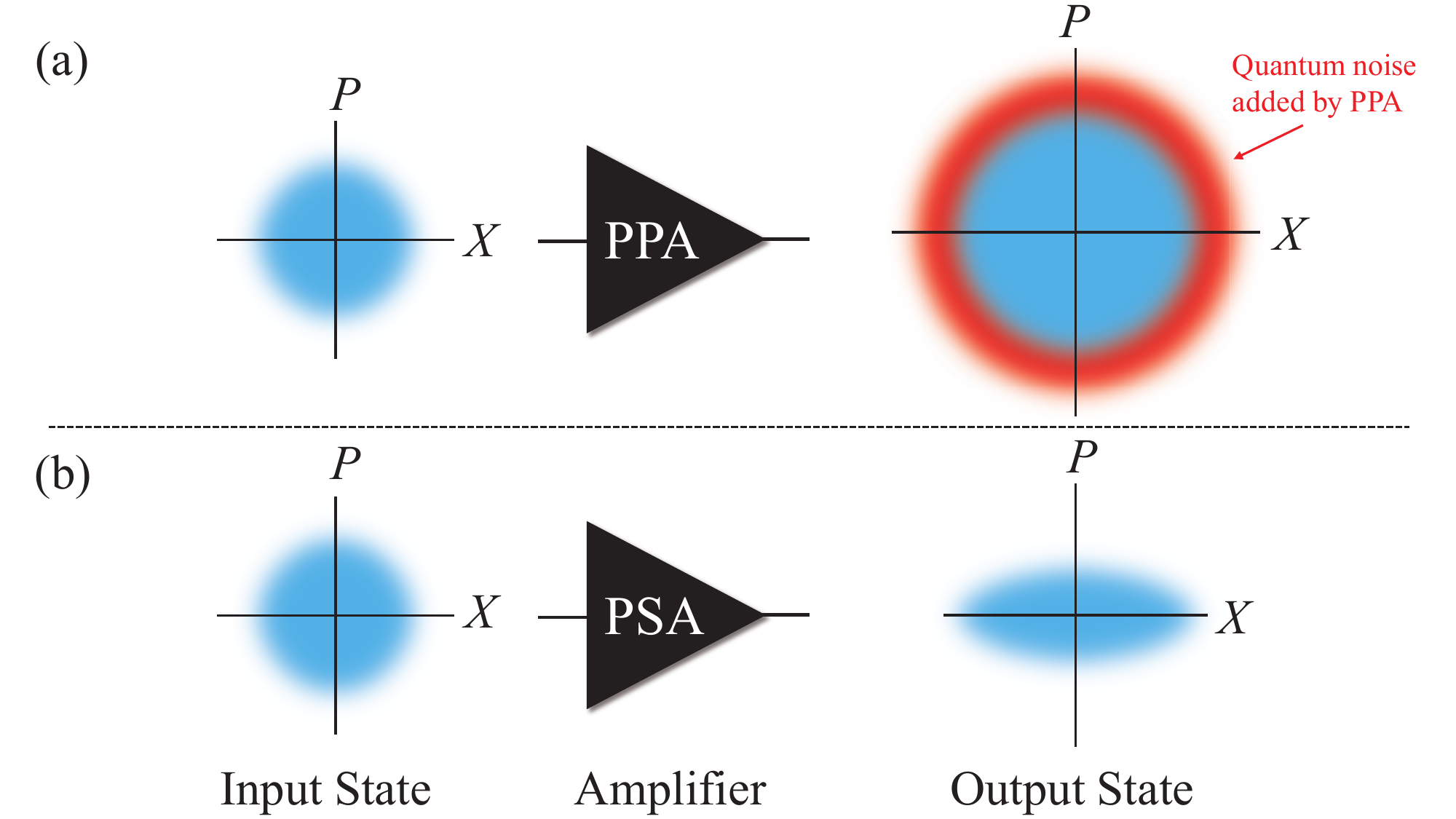}
{An illustration of the effects in phase space of the two forms of quantum-limited amplification on an input vacuum state. (a) Phase-preserving amplification (PPA) boosts both quadratures symmetrically and adds a non-zero amount of noise. (b) In contrast, with phase-sensitive amplification (PSA) one quadrature is boosted while the other is suppressed. Importantly, no added noise is introduced.\label{fig:Amplifiers}}

\begin{table*}[t!]
\centering
\resizebox{0.98\textwidth}{!}{%
\begin{tabular}{|l|c|c|c|}
\hline
\textbf{Pre-amplification} & \textbf{Entanglement} & \textbf{Non-classicality} & \textbf{Sub-vacuum conditional prediction} \\
\hline
None  
& $\left(\dfrac{\eta}{1-\eta}\right)\left[1-n_A(G_A-1) \right]$ 
&$\left(\dfrac{\eta}{1-\eta}\right)\left[1-n_A(G_A-1) \right]$ 
& $\frac{\eta }{1-\eta}
\left[
\frac{G_A }{2}\left(1-\frac{1}{\cosh 2r}\right)-(G_A-1) \left(n_A+\frac{1}{2}\right)
\right]-\frac{1}{2} $ \\
\hline
Phase-preserving 
& $\left(\dfrac{\eta}{1-\eta}\right)\left[1-n_A(G_A-1) \right]$ 
& $\left(\dfrac{\eta}{1-\eta}\right)\left[1-n_A(G_A-1) \right]$ 
& $\frac{\eta }{1-\eta}
\left[
\frac{G_A }{2}\left(1-\frac{G}{\cosh 2r}\right)-(G_A-1) \left(n_A+\frac{1}{2}\right)
\right]-\frac{1}{2} $ \\
\hline
Phase-sensitive 
& $\left(\dfrac{\eta}{1-\eta}\right)\left[1-n_A(G_A-1) \right]$ 
& $\left(\dfrac{\eta}{1-\eta}\right)\left[1+G_A\left(\frac{a^2-1}{2}\right)-n_A(G_A-1) \right]$ 
& $\frac{\eta }{1-\eta}
\left[
\frac{G_A a^2}{2}\left(1-\frac{1}{\cosh 2r}\right)-(G_A-1) \left(n_A+\frac{1}{2}\right)
\right]-\frac{1}{2} $ \\
\hline
\end{tabular}
}
\caption{Threshold mean noise photon number, $n_\text{ch}$ for entanglement, non-classicality, and sub-vacuum conditional prediction of the signal quadrature for a two-mode squeezed vacuum (TMSV) under one-sided thermal loss, with and without pre-amplification. The thresholds are defined in Sec. \ref{Threshold_Conditions_for_Quantum_Resources} and detailed calculations of each entry in the table are given in the Appendix.}
\label{threshold_formulas_table}
\end{table*}

Even after pre-amplification, the signal mode power is too small to be detectable at the receiver. Therefore, further conventional amplification is applied in two stages before transmission, as shown in Fig. \ref{fig:Microwave_Channel_Schematic}. Specifically, we assume the signal is amplified by a high-electron-mobility transistor (HEMT) amplifier at an intermediate temperature with gain $G_H$ and noise temperature $T_H$, and by a low-noise amplifier (LNA) at room temperature at the antenna feed with gain $G_L$ and noise temperature $T_L$. These stages are lumped together into a single stage of PPA with gain
$G_A = G_H G_L$ and noise temperature via the cascade equation \cite{pozar2021microwave}
\begin{align}
    T_A = T_\text{JPC}+\frac{G_H-1}{G_H}T_H +\frac{G_L-1}{G_H G_L}T_L,
    \label{amplifier_noise_temp}
\end{align}
where $T_\text{JPC}$ is the temperature of the input source (i.e., the JPC). In accord with Eq. \ref{PPA_transform}, this stage effects the transformation
\begin{align}
X\to \sqrt{G_A}X+X_\text{noise}, \;\;\; P\to \sqrt{G_A}P+P_\text{noise},
\label{HL_transform}
\end{align}
where $X_\text{noise}$ and $P_\text{noise}$ are quadratures of an effective thermal mode with mean photon number $n_A \approx k T_A/(hf)$, where $k$ is Boltzmann's constant, $h$ is Planck's constant, and $f$ is the frequency. The noise mode has quadrature variances 
\begin{align}
    \mathrm{Var}(X_\text{noise})=\mathrm{Var}(P_\text{noise})=(G_A-1)\left(n_A+\frac{1}{2}\right).
    \label{HL_noise_var}
\end{align}

We model all the effects of noise and loss in open-air, room-temperature transmission and reception with a Gaussian bosonic channel of transmissivity $\eta$ and noise variance $\nu = n_\text{ch}+1/2$, where $n_\text{ch}$ is the mean number of photons in the thermal noise mode. Given an input with quadratures $X$ and $P$, the action of the channel is the transformation 
\begin{align}
    X\to \sqrt{\eta}X+\sqrt{1-\eta}X_\text{ch},     \label{noise_loss_channel_X}\\
    P\to \sqrt{\eta}P+\sqrt{1-\eta}P_\text{ch},
    \label{noise_loss_channel_P}
\end{align}
where $X_\text{ch}$ and $P_\text{ch}$ are the quadrature operators of the noise mode. The values of $\eta$ and $n_\text{ch}$ are related to the parameters of a realistic microwave channel model in Sec. \ref{Microwave_Channel_Model}.

Within this model, we identify three thresholds for a fixed amount of loss ($\eta$) as noise ($n_\text{ch}$) increases from zero. These threshold noise levels, defined in detail in Sec. \ref{Threshold_Conditions_for_Quantum_Resources}, mark the loss of (1) entanglement, (2) non-classicality, and (3) sub-vacuum conditional prediction of the signal quadrature. The threshold conditions are collected in Table \ref{threshold_formulas_table}. The details of this table are given in the appendix, but some conclusions can be drawn immediately. First, neither form of pre-amplification extends the threshold for entanglement beyond that of the original TMSV state. In other words, these forms of pre-amplification do not protect entanglement from the noise and loss in our model\cite{agarwal2010much}. Second, the thresholds for non-classicality and sub-vacuum conditional prediction can, in fact, be extended, but only by PSA. 

These conclusions are clarified by examining a single set of parameters. Figure \ref{fig:All_thresholds_plot} shows the non-zero thresholds for non-classicality and sub-vacuum conditional prediction as functions of the PSA pre-amplifier power gain, $a^2$, for parameters derived from the microwave channel model of Sec. \ref{Microwave_Channel_Model} that assumes room-temperature transmit and receive horn antennas separated by 3 meters. The reported thresholds are interpreted as the maximum equivalent thermal occupation of the modeled channel for which the specified resource survives. Comparison with the approximately 300-K thermal occupation therefore measures robustness margin within the model. The threshold noise level for entanglement is zero with or without pre-amplification. Entanglement cannot survive open-air transmission in this scenario. Similarly, the thresholds for the other two resources with PPA pre-amplification are zero. But notably, the thresholds for non-classicality and sub-vacuum conditional prediction are much higher with PSA pre-amplification. In this case, the threshold mean noise photon number for sub-vacuum conditional prediction is up to two orders of magnitude above the thermal background value of approximately 1250 photons over a wide range of physically realizable gains. The threshold for non-classicality is nearly another order of magnitude larger. It seems to be the case that these resources, protected by PSA, offer the best chances for successful sensing or communication protocols through open-air, room-temperature microwaves.

The rest of the paper is organized as follows. Section \ref{Threshold_Conditions_for_Quantum_Resources} defines the threshold conditions for each of the three quantum resources. Section \ref{Microwave_Channel_Model} relates the parameters of a noise-loss bosonic channel to those of an open-air, room-temperature microwave channel. Section \ref{Nonideal_Effects} adds non-ideal PSA effects of phase misalignment and added noise to the model. In Sec. \ref{Conclusion} we draw conclusions. The details of applying the threshold conditions from Sec. \ref{Threshold_Conditions_for_Quantum_Resources} are given in the Appendix.

\section{Threshold Conditions for Quantum Resources}
\label{Threshold_Conditions_for_Quantum_Resources}
Here we determine the conditions that define the thresholds, collected in Table \ref{threshold_formulas_table}, for the three quantum resources of interest. Each threshold is a function of the channel parameters $n_\text{ch}$ and $\eta$, the squeezing parameter $r$ that characterizes the TMSV state, and the gains $G$, $a$, and $G_A$ applied in each case. In some cases, the threshold is independent of pre-amplifier gain ($G$ or $a$), indicating that amplification provides no benefit. In other cases, the threshold does depend on the amplifier gain, indicating the effectiveness of pre-amplification to push a threshold to a higher or lower level of channel noise. 

To reiterate, the resources of interest are entanglement, Glauber–Sudarshan non-classicality, and sub-vacuum conditional prediction of the input-referred signal quadrature. The first two are properties of the output two-mode state, whereas the third is an operational property of the source–channel system with retained idler.
\Figure[t!](topskip=0pt, botskip=0pt, midskip=0pt)[width=0.95\columnwidth]{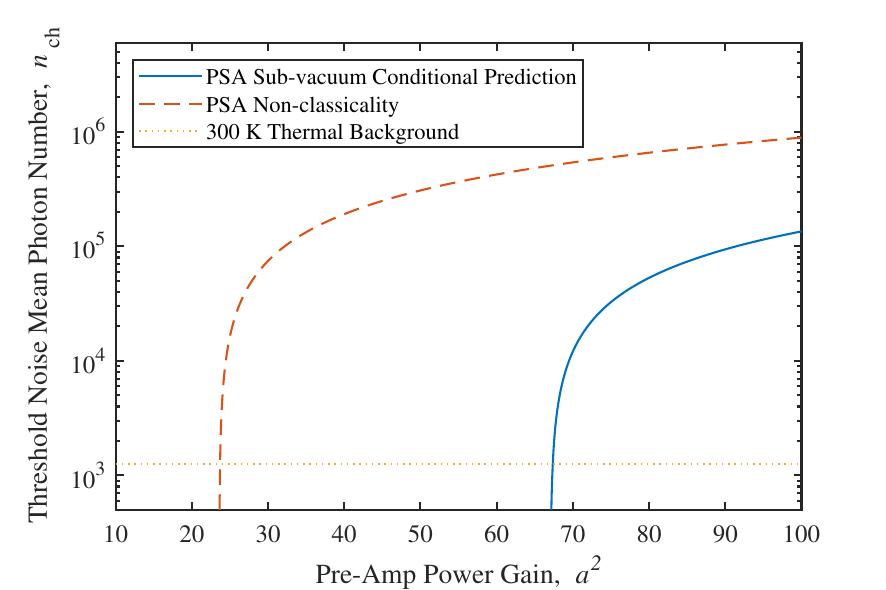}
{Threshold mean noise photon number, $n_\text{ch}$, as a function of pre-amplification gain for PSA for non-classicality (red, dashed line) and sub-vacuum conditional prediction (blue, solid line). The parameters are $\eta = 0.023$, $G_A = 10^6$, and $n_A = 11.3$.\label{fig:All_thresholds_plot}}

\subsection{Entanglement}
A two-mode, Gaussian state is entangled if the smallest symplectic eigenvalue of the partially transposed covariance matrix satisfies \cite{weedbrook2012gaussian}
\begin{align}
    \tilde{\nu}_- < 1/2.
    \label{entanglement_threshold}
\end{align}
Such a state exhibits non-separable correlations, i.e., correlations that cannot be created by local operations alone. Entanglement is the resource utilized by many canonical quantum protocols, including teleportation, entanglement swapping, purification, and some forms of quantum key distribution. 

The entanglement thresholds for the TMSV after passing the signal mode through a noise-loss channel with and without pre-amplification are shown in the second column of Tab. \ref{threshold_formulas_table}. The threshold is the same in all three cases. Notably, a very small amount of noise ($n_A$) and gain ($G_A$) from the HEMT/LNA cascaded amplifiers pushes the threshold to zero, meaning entanglement will certainly not survive through the open-air channel. The common threshold is independent of the pre-amplifier gains, $G$ or $a$. Thus, within the local Gaussian PPA and PSA pre-amplification strategies considered here, increasing the pre-amplifier gain does not extend the entanglement threshold. Interestingly, the threshold is also independent of the squeezing parameter $r$, so additional squeezing provides no additional robustness to noise and loss. 

\subsection{Non-classicality}
A two-mode Gaussian state is non-classical if
\begin{align}
    \textbf{V}-\frac{1}{2}\textbf{I}\not\ge 0,
    \label{nonclass_condition}
\end{align}
meaning that the matrix is not positive semi-definite, or, equivalently, the matrix has at least one negative eigenvalue. For such a state, the Glauber-Sudarshan $P$ function is not a proper probability distribution. There exists a quadrature or combination of quadratures with variance below vacuum \cite{vogel2000nonclassical}, but correlations may be classical \cite{korolkova2019quantum}. A non-classical Gaussian state need not be entangled; its non-classicality may instead arise from local or collective squeezing.

The non-classicality thresholds for the TMSV after passing the signal mode through a noise-loss channel with and without pre-amplification are shown in the third column of Table \ref{threshold_formulas_table}. The non-classicality threshold without amplification is the same as the entanglement threshold. Both properties are lost at the same noise level, which is effectively zero for realistic amounts of HEMT/LNA noise ($n_A$) and gain ($G_A$). This condition also holds when the signal mode is pre-amplified with PPA. Notably, the threshold with PSA increases with the power gain $a^2$. The red, dashed line in Figure \ref{fig:All_thresholds_plot} shows how PSA can raise the noise threshold orders of magnitude above the thermal background. Thus PSA can plausibly preserve this quantum resource through an open-air, room-temperature channel.  This finding is one of the major results of this paper.

\subsection{Sub-vacuum conditional prediction of the signal quadrature}
The third quantum resource we consider differs from the previous two in that it is not a property of the two-mode state after open-air transmission. Rather, it is an operational resource of the source-channel-idler system as a whole. Specifically, it is the capability to predict the amplified and transmitted  input-referred signal mode quadrature with sub-vacuum minimum mean-square error conditioned on the measured idler quadrature. The measured idler can  provide predictive power about the signal beyond vacuum noise. Importantly, this capability is due to residual correlations remaining from the entanglement of the original TMSV. Such sub-vacuum conditional prediction cannot be reproduced by replacing the TMSV with a classical state described by a nonnegative Glauber-Sudarshan $P$ distribution. For this reason, we refer to this capability as a quantum resource.

As shown in Eq. \ref{TMSV_cond_pre}, the initial TMSV exhibits sub-vacuum conditional prediction. Let \(X_s^{\rm out}\) denote the measured signal quadrature at the receiver. The measured quadrature is referred to the input by rescaling measurement outcomes from this mode by a factor of $1/\sqrt{\eta G G_A}$. In this expression, $G$ is the power gain of PPA. But this definition includes the case of no pre-amplification  by setting $G=1$, and the case of PSA by replacing $G$ with $a^2$. The input-referred measurement outcomes are equivalent to measuring a fictitious mode with quadrature operator
\begin{align}
\tilde X_s=\frac{X_s^\text{out}}{\sqrt{\eta G G_A}}.
\end{align}
Note that input referencing is not itself another stage of amplification. Rather,  it is classical post-processing that rescales signal and noise equally. In the appendix, it is shown that
\begin{align}
    \tilde X_s=X_s + Z,
\end{align}
where $Z$ represents the contribution of all noise sources (amplifiers, background, etc.) Specific functional forms for $Z$ are given in the appendix for the cases of PPA and PSA pre-amplification. We assume that \(Z\) is statistically independent of the original TMSV quadratures \((X_s,X_i)\). Consequently,
\begin{align}
    \operatorname{Cov}(Z,X_i)=\operatorname{Cov}(Z,X_s)=0.
\end{align}
It follows that
\begin{align}
    \operatorname{MMSE}(\tilde X_s|X_i)=
\frac{1}{2\cosh 2r}
+
\operatorname{Var}(Z).
\end{align}
Sub-vacuum conditional prediction survives as long as the added input-referred noise is sufficiently small that
\begin{align}
    \operatorname{Var}(Z)<\frac{1}{2} -\frac{1}{2\cosh 2r}. 
\end{align}
This condition defines the threshold for this quantum resource. Unlike Eqs. \ref{entanglement_threshold}
 and \ref{nonclass_condition}, this is not simply a function of the covariance matrix of the output two-mode state. It depends on the channel and amplifier parameters, as well as the signal-idler state. This is the sense in which sub-vacuum conditional prediction is an operational resource of the source-channel-idler system as a whole.

The fourth column of  Table \ref{threshold_formulas_table}  shows the noise thresholds below which $\operatorname{MMSE}(\tilde X_s|X_i)<1/2$. As was the case with the previous quantum resources, without pre-amplification the noise threshold is effectively zero for realistic amounts of HEMT/LNA noise ($n_A$) and gain ($G_A$). Interestingly, PPA actually lowers the noise threshold as $G$ increases due to the noise it adds even when quantum limited. Strikingly, PSA can raise the threshold significantly. In fact, the threshold increases with the power gain $a^2$.

The blue solid line in Fig. \ref{fig:All_thresholds_plot} shows how PSA with sufficient gain $a^2$ can raise the noise threshold more than an order of magnitude above the blackbody, thermal background. Thus, this form of quantum-limited pre-amplification can plausibly preserve this quantum resource through an open-air, room-temperature channel.  This finding is one of the major results of this paper. 

Also notably, the blue line is always below the red dashed line in Fig.  \ref{fig:All_thresholds_plot}. In fact, subtracting column 3 from column 2 in row 3 of Tab. \ref{threshold_formulas_table} gives
\begin{align}
\frac{\eta}{1-\eta}
\left[
\frac12+\frac{G_Aa^2}{2\cosh 2r}
\right]
+\frac12 >0
,
\end{align}
implying that sub-vacuum conditional prediction only occurs within a subset of the parameter regime in which the output state remains Glauber–Sudarshan nonclassical. This result supports the characterization of sub-vacuum conditional prediction as a quantum resource, albeit a resource of the complete system, not just the microwave field state.

As an independent numerical check of the expressions in Tab. \ref{threshold_formulas_table}, we propagated the covariance matrix directly through the amplifier and thermal-loss transformations and evaluated the PPT, Glauber–Sudarshan nonclassicality, and conditional-MMSE criteria numerically for the representative parameter set. The resulting threshold crossings agree with the analytical expressions in Tab. \ref{threshold_formulas_table} to numerical precision.

The threshold calculations assume that the effective channel transmissivity and amplifier gains are calibrated. Slowly varying gain or transmission errors can be incorporated by estimating these parameters from the measured data, while faster fluctuations act as an additional noise source and therefore reduce the available margin below the ideal thresholds. Likewise, the optimum conditional gain $g_{\rm opt}$ can be obtained experimentally via Eq. \ref{opt_gain} from the measured covariance $\operatorname{Cov}(X_s,X_i)$ and variance $\operatorname{Var}(X_i)$, rather than relying solely on nominal amplifier settings.

\section{Microwave Channel Model}
\label{Microwave_Channel_Model}
Here we describe the model used to choose the parameters behind the results in Fig. \ref{fig:All_thresholds_plot}. As depicted in Fig. \ref{fig:Microwave_Channel_Schematic}, a Josephson parametric converter is used to generate the entangled signal and idler modes inside a dilution refrigerator, providing an environment with a temperature on the order of ten millikelvin. The signal mode is pre-amplified by a second Josephson junction circuit (JPA), and amplified again by the cascade of a HEMT amplifier at an intermediate temperature and a LNA at room temperature. The signal is then transmitted through a horn antenna at room temperature, propagates a short distance through open air, and is received by a second horn antenna and measured. 

The quantum channel  of Eqs. \ref{noise_loss_channel_X} and \ref{noise_loss_channel_P} contains two parameters, $\eta$ and $n_\text{ch}$, which quantify the total loss and added noise, respectively, from transmission to reception. The channel transmissivity $\eta$ is the product $\eta = \eta_\text{tx} \eta_\text{prop}$. Here $\eta_\text{tx}$ accounts for dissipative losses in the warm transmit chain between the output of the LNA and the antenna feed. Antenna mismatch is assumed negligible. The propagation factor $\eta_\text{prop}$ is determined by the Friis power transmission equation \cite{pozar2021microwave}
\begin{align}
    \eta_\text{prop}=G_t G_r \left(\frac{\lambda}{4\pi R}\right)^2,
\end{align}
where $G_t$ and $G_r$ are the transmit and receive antenna gains, respectively, $\lambda$ is the wavelength, and $R$ is the distance between antennas. The gains include antenna radiation efficiency, so radiation losses are not included separately in $\eta_\text{tx}$.

Next we determine the mean noise photon number $n_\text{ch}$. We assume the transmit antenna is a room-temperature metal horn fed from the output of the LNA through a warm transmit chain with transmissivity  $\eta_\text{tx}$. This parameter represents dissipative feed and interconnect losses preceding the antenna; antenna radiation efficiency is already included in $G_t$. A simple model \cite{balanis2015antenna} is
\begin{align}
    T_\text{tx} =\eta_\text{tx} T_\text{A} + (1-\eta_\text{tx}) T_\text{phys},
\end{align}
where $T_\text{A}$ is given in Eq. \ref{amplifier_noise_temp}, and $T_\text{phys}$ is the physical temperature of the lossy hardware. 
The effective antenna temperature $T_\text{ant}$ is then
\begin{align}
    T_\text{ant}=F T_\text{tx} +(1-F) T_\text{phys}
\end{align}
where $F$ is the fraction of the receive beam occupied by the transmit horn. A rough  geometric estimate is
\begin{align}
F \approx \frac{\Omega_t}{\Omega_B},
\end{align}
where $\Omega_t$ is the solid angle subtended by the transmit horn and $\Omega_B$ is the receive-beam solid angle. For an aligned far-field link,
\begin{align}
\Omega_t \approx \frac{A_t}{R^2},
\qquad
\Omega_B \approx \frac{4\pi}{D_r},
\end{align}
so
\begin{align}
F \approx \frac{A_t D_r}{4\pi R^2},
\end{align} where $D_r$ is the directivity of the receive-antenna and $A_t$ is the physical aperture area of the transmit antenna. For representative 20-dBi horn dimensions at a 3-m separation, this gives $F\sim0.03$. Here $A_t/R^2$ is the solid angle subtended by the transmit aperture in the small-angle approximation, and $D_r$ is dimensionless, so that $F$ is dimensionless as required. Finally, the mean noise photon number in the received mode is determined by
\begin{align}
    n_\text{ch} \approx \frac{kT_\text{ant}}{h f},
\end{align}
where $k$ is Boltzmann's constant, $h$ is Planck's constant, and $f$ is the frequency.

The model parameters are collected in Table \ref{model_values_table}. Plausible ranges for each parameter are displayed along with the parameter set used to generate Fig. \ref{fig:All_thresholds_plot}. The assumed pre-amplifier frequency and gain ranges are consistent with experimentally demonstrated Josephson parametric amplifiers, which have achieved near-quantum-limited operation with approximately 20 dB gain across much of the 4–12 GHz band \cite{sivak2020josephson}.

\begin{table}[t!]
\centering
\setlength{\tabcolsep}{5pt}
\renewcommand{\arraystretch}{1.5}
\resizebox{0.99\columnwidth}{!}{%
\begin{tabular}{|c|c|c|c|}
\hline
\textbf{Parameter} & \textbf{Symbol} & \textbf{Plausible Range} & \textbf{Example Value} \\
\hline
Frequency & $f$ & $4-12$ GHz (C to X band)& $5$ GHz  \\
\hline
Wavelength & $\lambda$ & $2.5-7.5$ cm & $6$ cm  \\
\hline
Link distance & $R$ & $2$--$5$ m & 3 m  \\
\hline
Squeezing parameter & $r$ & $0.3$--$1.0$ & 0.5  \\
\hline
PPA power gain& $G$ & $10$--$100$ & 100 \\
\hline
PSA amplitude gain & $a$ & $3.2$--$10$ & 10 \\
\hline
JPC temperature & $T_\text{JPC}$ & $0.05$--$1$ K &$0.1$ K  \\
\hline
Warm hardware temperature  & $T_{\text{phys}}$ & $290-300$ K & $300$ K  \\
\hline
Warm transmit-chain efficiency & $\eta_\text{tx}$ & $0.8$--$0.99$ & 0.90  \\
\hline
Transmit antenna gain & $G_t$ & $10$--$20$ dBi & 20 dBi  \\
\hline
Receive antenna gain & $G_r$ & $10$--$20$ dBi & 20 dBi  \\
\hline
HEMT noise temperature & $T_H$ & $2$--$10$ K & 4 K    \\
\hline
HEMT gain & $G_H$ & $10$--$30$ dB & 30 dB    \\
\hline
LNA noise temperature & $T_L$ & $50$--$200$ K & 120 K    \\
\hline
LNA gain & $G_L$ & $15$--$35$ dB & 30 dB    \\
\hline
Beam filling factor & $F$ & 0.02-0.05 & 0.03  \\
\hline
\end{tabular}}
\caption{Representative parameters for a microwave two-mode squeezing transmission experiment with phase-sensitive pre-amplification. Antenna gains include radiation efficiency; impedance mismatch is neglected. The Example Value column gives the parameter set used to generate Fig. \ref{fig:All_thresholds_plot}.}
\label{model_values_table}
\end{table}

\section{Non-ideal Phase Sensitive Amplification}
\label{Nonideal_Effects}
The analysis so far has assumed ideal, quantum-limited performance from the pre-amplifier. Since PSA has been shown to preserve non-classicality and sub-vacuum conditional prediction through the noise and loss of an open-air channel model, it is important to know how robust these results are to two important non-ideal effects: phase-misalignment and PSA-added noise. Here we examine how each effect changes the thresholds in Tab. \ref{threshold_formulas_table}. We find that phase accuracy is relatively forgiving, whereas added noise is the more critical requirement.

\subsection{Phase misalignment}
A phase difference $\phi$ between the PSA pre-amplification and the homodyne measurements has a very different effect on the threshold for non-classicality versus sub-vacuum conditional prediction. With respect to the former, PSA phase misalignment does not change the threshold at all. The reason is that a phase error merely rotates the squeezing ellipse. The PSA still produces the same amount of squeezing or nonclassicality; it is simply rotated in phase space. The threshold actually does not depend on the measurements made on the idler and received mode.

For sub-vacuum conditional prediction, a resource of the entire source-channel-idler system, phase misalignment does affect the threshold. A phase difference $\phi$ between the PSA pre-amplification and the homodyne measurements can be accounted for by replacing the ideal PSA of Eq. \ref{PSA_transform} with an amplified output quadrature of the form
\begin{align}
X_s'
=
g_{\phi}X_s+h_{\phi}P_s,
\end{align}
where 
\begin{align}
g_{\phi}=a\cos^2\phi+a^{-1}\sin^2\phi,
\end{align}
and
\begin{align}
h_{\phi}=\left(a-a^{-1}\right)\sin\phi\cos\phi.
\end{align}
If we refer the received quadrature back to the input using the actual $X_s$-gain $g_{\phi}$, then
\begin{align}
\widetilde X_s
=
X_s+\frac{h_{\phi}}{g_{\phi}}P_s+Z.
\end{align}
For a TMSV,
\begin{align}
\operatorname{Var}(P_s)=\frac{c}{2},
\qquad c=\cosh 2r,
\end{align}
and \(P_s\) is uncorrelated with the measured idler quadrature \(X_i\). Therefore the orthogonal-quadrature leakage caused by phase error adds variance
\begin{align}
N_\phi=
\frac{c}{2}
\left[\frac{h_{\phi}}{g_{\phi}}\right]^2
\end{align}
to the conditional prediction error. Adding this additional noise to $Z_{PSA}$ in the derivation of App. \ref{App_PSA} leads to the modified threshold for sub-vacuum conditional prediction

\begin{align}
&
\frac{\eta}{1-\eta}
\Bigg\{
\frac{G_A}{2}
\left[
g_{\phi}^2\left(1-\frac{1}{c}\right)-c h_{\phi}^2
\right]\nonumber
\\
&\qquad
-(G_A-1)
\left(n_A+\frac12\right)
\Bigg\}
-\frac12 .
\end{align}

\Figure[t!](topskip=0pt, botskip=0pt, midskip=0pt)[width=0.95\columnwidth]{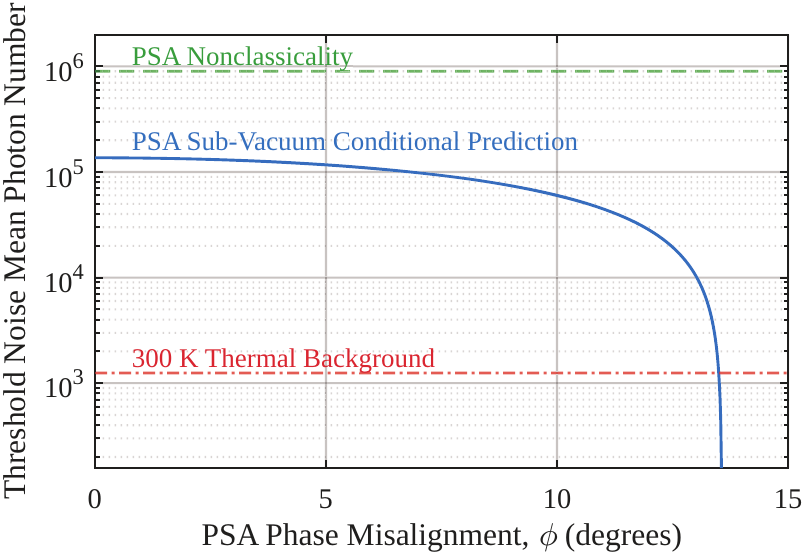}
{\parbox[t]{\columnwidth}{Threshold mean noise photon number  for non-classicality (green dashed line) and for sub-vacuum conditional prediction (solid blue line) as a function of phase misalignment $\phi$ between the PSA-preamplifier and the homodyne measurements. For a sense of scale, the thermal background mean photon number is also indicated (red dashed-dotted line).}\label{fig:Phase_misalignment}}

\subsection{PSA added noise}

A second important nonideality is excess noise added by the PSA. Although Josephson parametric amplifiers can operate near the quantum limit, experimentally measured added noise can exceed this limit under some operating conditions \cite{bergeal2010phase, sivak2020josephson}. We model this phenomenologically by replacing the ideal amplified-quadrature transformation \(X_s\rightarrow aX_s\) with
\begin{align}
X_s\rightarrow a(X_s+X_{\rm PSA}),
\end{align}
where \(X_{\rm PSA}\) is independent, zero-mean Gaussian noise with input-referred quadrature variance \(\operatorname{Var}(X_{\rm PSA})=N_{\rm PSA}\). Referring the received quadrature back to the input therefore adds \(N_{\rm PSA}\) directly to the conditional prediction error. The threshold for sub-vacuum conditional prediction becomes
\begin{equation}
\begin{aligned}
&
\frac{\eta}{1-\eta}
\Bigg[
G_Aa^2
\left(
\frac{1}{2}\left(1-\frac{1}{c}\right)
-N_{\rm PSA}
\right)
\\
&\qquad
-(G_A-1)
\left(n_A+\frac{1}{2}\right)
\Bigg]
-\frac{1}{2},
\end{aligned}
\end{equation}
where \(c=\cosh2r\). Similarly, the Glauber--Sudarshan nonclassicality threshold becomes
\begin{equation}
\begin{aligned}
&
\frac{\eta}{1-\eta}
\Bigg[
1+\frac{G_A(a^2-1)}{2}
\\
&\qquad
-G_Aa^2N_{\rm PSA}
-n_A(G_A-1)
\Bigg].
\end{aligned}
\end{equation}
Thus, PSA-added noise lowers both thresholds linearly at the same rate. For the representative parameters of Table \ref{model_values_table}, the sub-vacuum conditional-prediction threshold remains above the 300 K thermal background for \(N_{\rm PSA}\lesssim0.057\), while nonclassicality remains above the thermal background up to \(N_{\rm PSA}\lesssim0.38\). For comparison, if the excess noise were isotropic Gaussian noise with equal variance in both quadratures, \(N_{\rm PSA}\) would equal the corresponding input-referred added mean photon number.

\Figure[t!](topskip=0pt, botskip=0pt, midskip=0pt)[width=0.95\columnwidth]{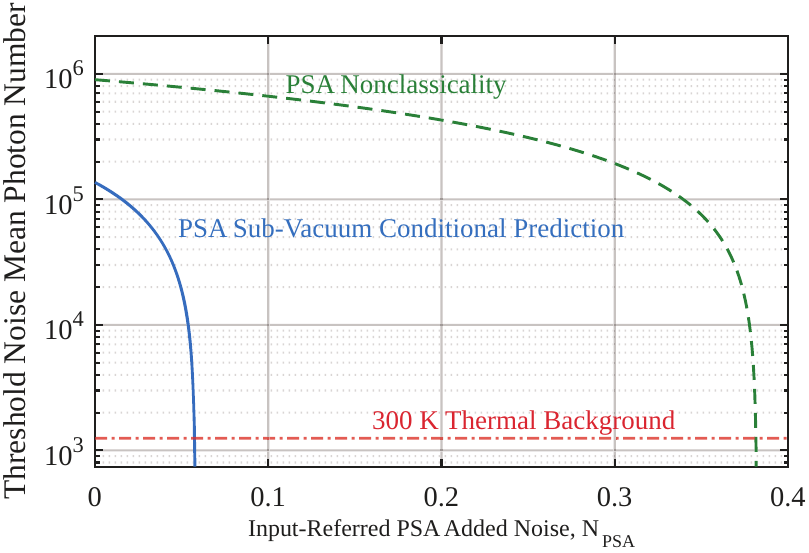}
{\parbox[t]{\columnwidth}{Threshold mean noise photon number as a function of variance of noise added by a non-ideal PSA for nonclassicality (green dashed line) and sub-vacuum conditional prediction (blue solid line). For a sense of scale, the thermal background mean photon number is also indicated (red dashed-dotted line).}\label{fig:PSA_noise}}

\section{Conclusion}
\label{Conclusion}
In this paper, we have determined that pre-amplification preserves certain quantum resources through an open-air, room-temperature microwave channel. Entanglement was shown to be the least robust resource, unable to withstand the modeled loss and noise without regard to pre-amplification. In part, this is due to the noise added by the HEMT and LNA. But even without these amplifiers, the threshold channel noise level is too low to be reached in a real experiment. In contrast, pre-amplification can protect non-classicality and sub-vacuum conditional prediction at noise levels orders of magnitude above the thermal occupancy, but only when the pre-amplification is phase sensitive.  

The model parameters here represent a 3-meter link between horn antennas with a frequency of 5 GHz. The margin of the noise thresholds above the thermal background is large enough to expect that longer distance links are achievable. The large margins may also allow our conclusions to survive effects not modeled here, such as polarization effects or multipath. Further study is needed to confirm this possibility.

Sub-vacuum conditional prediction may enable a variety of technological applications because measurements of the retained idler provide predictive power about measurements of the signal mode beyond the vacuum limit. Such correlations are impossible for modes originating from classical states. Possible applications of this quantum resource in the microwave regime include microwave metrology \cite{caves1981quantum,walls1983squeezed,engen1992microwave,giovannetti2011advances}, remote sensing \cite{barzanjeh2015microwave,luong2019receiver}, conditional noise cancellation \cite{steinlechner2013quantum}, and distributed sensing with correlated receiver arrays \cite{proctor2018multiparameter, eldredge2018optimal}.

\appendix
\section{Details of Threshold Calculations}
\label{Details_of_Threshold_Calculations}
Here we present the details of the derivation of each threshold shown in Table \ref{threshold_formulas_table}. The thresholds for the case with no pre-amplification follow from the thresholds for PPA when the gain $G$ is set to unity.
The covariance matrices in each case will be of the form 
\begin{align}
\textbf{V}&=\frac{1}{2}
\left ( \begin{array}{cc}
     \textbf{A} &  \textbf{C} \\
    \textbf{C}^T & \textbf{B} 
\end{array}\right),
\label{cov_form}
\end{align}
where $\textbf{A}$, $\textbf{B}$, and $\textbf{C}$ are two-dimensional matrices.

\subsection{Threshold Calculations for TMSV with Phase-Preserving Amplification}
A TMSV state has covariance matrix where
\begin{align}
\textbf{A} = \textbf{B} =\frac{1}{2}c\textbf{I}_2,\; \textbf{C}=\frac{1}{2}s\textbf{\sigma_Z},
\label{cov_TMSV}
\end{align}
where $\textbf{I}_2$ is a two-dimensional identity matrix and $\textbf{\sigma_Z}$ is a Pauli matrix.
Here $c=2N+1=\cosh{2r}$ and $s=2\sqrt{N(N+1)}=\sinh{2r}$, where $N$ is the mean photon number and $r$ is the squeezing parameter. Applying quantum-limited phase-preserving amplification with gain $G$, as embodied in the transformation of Eq. \ref{PPA_transform}, to the TMSV covariance matrix of Eq. \ref{cov_TMSV}, then applying an additional stage of PPA with gain $G_A$ and noise temperature $T_A$, and, finally, passing the signal mode through a noise-loss channel that enacts the transformation of Eqs. \ref{noise_loss_channel_X} and \ref{noise_loss_channel_P}, gives an output covariance matrix where
\begin{align}
\textbf{A} =\frac{1}{2}\left[\eta c'+2(1-\eta)\nu\right]\textbf{I}_2,\;  \nonumber \\
\textbf{B} =\frac{1}{2}c\textbf{I}_2, \;\textbf{C}=\frac{1}{2}\sqrt{\eta}s'\textbf{\sigma_Z},
\label{cov_TMSV_PPA_HL_noise}
\end{align}
with $c'\equiv \left[ GG_A c + G_A(G-1)+(G_A-1)(2n_A+1)\right]$ and $s'\equiv \sqrt{GG_A}s$.

For a covariance matrix of the form 
of Eq. \ref{cov_form},
the smallest symplectic eigenvalue of the partially transposed covariance matrix is 
\begin{align}
    \tilde{\nu}_- =\sqrt{\frac{\tilde{\Delta}-\sqrt{\tilde{\Delta}^2-4\text{ det}\textbf{V}}}{2}},
\end{align}
with 
\begin{align}
    \tilde{\Delta}\equiv \text{ det} \textbf{A} + \text{ det} \textbf{B} - 2\text{ det} \textbf{C}.
    \label{delta_tilde}
\end{align}
The entanglement threshold of Eq. \ref{entanglement_threshold} can be squared and rearranged to give the condition
\begin{align}
    \tilde{\Delta}-4\text{ det}\textbf{V}-\frac{1}{4} > 0.
    \label{delta_condition}
\end{align}The condition for entanglement in the form of Eq. \ref{delta_condition} becomes
\begin{align}
    \left[w^2 -\left(u-\frac{1}{2}\right)\left(v-\frac{1}{2}\right)\right]\times &\nonumber \\
    \left[\left(u+\frac{1}{2}\right)\left(v+\frac{1}{2}\right)-w^2\right]&>0,
    \label{ent_PPA_raw_condition}
\end{align}
where $u \equiv \left[\eta c'+2(1-\eta)\nu\right]/2$, $v\equiv c/2$, and $w\equiv  \sqrt{\eta}s'/2$. The positive semi-definiteness of the covariance matrix requires the bracketed term on the right to be positive. 
Then the entanglement condition is equivalent to the condition for positivity of the bracketed term on the left, and the result the first column, second row of Table \ref{threshold_formulas_table} follows. 

To derive the condition for non-classicality, a similarity transformation applied to $\textbf{V}-\textbf{I}/2$ gives the matrix
\begin{align}
\left ( \begin{array}{cccc}
     u-\frac{1}{2} &  w & 0 & 0\\
    w &v-\frac{1}{2} & 0 & 0\\
    0 & 0 & u-\frac{1}{2}& -w \\
     0 & 0 & -w & v-\frac{1}{2}
\end{array}\right),
\label{cov_TMSV_reorder}
\end{align}
which has the same eigenvalues as $\textbf{V}-\textbf{I}/2$. The block structure of this matrix implies that the eigenvalues of $\textbf{V}-\textbf{I}/2$ are just the eigenvalues of the two non-zero blocks. A $2 \times 2$ matrix is positive semi-definite if and only if both diagonal elements are nonnegative and the determinant is nonnegative. The determinant gives the condition
\begin{align}
    w^2 -\left(u-\frac{1}{2}\right)\left(v-\frac{1}{2}\right)>0,
\end{align}
which is the same as the entanglement condition Eq. \ref{ent_PPA_raw_condition}, hence the result in Table \ref{threshold_formulas_table}.

To derive the threshold for sub-vacuum conditional prediction with phase-preserving pre-amplification, begin with the TMSV signal quadrature \(X_s\), apply the phase-preserving transformations of Eqs. \ref{PPA_transform} and \ref{HL_transform}, and then apply the thermal-loss channel of Eqs. \ref{noise_loss_channel_X} and \ref{noise_loss_channel_P}. The signal quadrature at the receiver is
\begin{align}
&X_s^{\rm out}=\nonumber \\
&
\sqrt{\eta G G_A}\,X_s
+
\sqrt{\eta G_A}\,X_{\rm add}
+
\sqrt{\eta}\,X_{\rm noise}
+
\sqrt{1-\eta}\,X_{\rm th},
\end{align}
where
\begin{align}
\operatorname{Var}(X_{\rm add})=\frac{G-1}{2},
\end{align}
\begin{align}
\operatorname{Var}(X_{\rm noise})
=
(G_A-1)\left(n_A+\frac12\right),
\end{align}
and
\begin{align}
\operatorname{Var}(X_{\rm th})
=
n_{\rm ch}+\frac12 .
\end{align}

Referring the measured quadrature to the input gives
\begin{align}
\widetilde X_s
=
\frac{X_s^{\rm out}}
{\sqrt{\eta G G_A}}
=
X_s+Z_{\rm PPA},
\end{align}
where
\begin{align}
Z_{\rm PPA}
=
\frac{1}{\sqrt G}X_{\rm add}
+
\frac{1}{\sqrt{G G_A}}X_{\rm noise}
+
\sqrt{\frac{1-\eta}{\eta G G_A}}X_{\rm th}.
\end{align}
The added-noise modes are statistically independent of the original TMSV signal and idler. Hence
\begin{align}
\operatorname{MMSE}(\widetilde X_s|X_i)
=
\operatorname{MMSE}(X_s|X_i)
+
\operatorname{Var}(Z_{\rm PPA}).
\end{align}
Sub-vacuum conditional prediction requires
\begin{align}
\operatorname{MMSE}(\widetilde X_s|X_i)<\frac12 .
\end{align}
Solving this inequality for the channel thermal occupation gives the phase-preserving threshold reported in Table \ref{threshold_formulas_table}. The case of no pre-amplification follows by setting \(G=1\).

\subsection{Threshold Calculations for TMSV with Phase-Sensitive Amplification}
\label{App_PSA}
Applying quantum-limited PSA with power gain $a^2$, as embodied in the transformation of Eq. \ref{PSA_transform}, to the TMSV covariance matrix of Eq. \ref{cov_TMSV}, then applying an additional stage of PPA with gain $G_A$ and noise temperature $T_A$, and, finally, passing the signal mode through a noise-loss channel that enacts the transformation of Eqs. \ref{noise_loss_channel_X} and \ref{noise_loss_channel_P}, gives an output covariance matrix where
\begin{align}
\textbf{A} &=\left ( \begin{array}{cc}
     A_X &  0 \\
    0 & A_P 
\end{array}\right),\nonumber \\
&\;  \textbf{B} =\frac{1}{2}c\textbf{I}_2, \;\textbf{C}=\frac{1}{2}\sqrt{q}s\left ( \begin{array}{cc}
     a &  0 \\
    0 & -a^{-1} 
\end{array}\right),
\label{cov_TMSV_PSA_HL_noise}
\end{align}
with $A_X \equiv m+\frac{ a^2q}{2}c$ and $A_P \equiv m+\frac{q}{2a^2}c$,  $m = \eta(G_A-1)\left(n_A+1/2\right)+ (1-\eta)\nu$, and $q \equiv \eta G_A$.

The condition for entanglement in the form of Eq. \ref{delta_condition} becomes
\begin{align}
    \tilde{\Delta}-&4\text{ det}\textbf{V}-\frac{1}{4}=\nonumber \\
    &\frac{1}{4}\left(c^2-1\right)(q+1+2m)(q+1-2m)>0.
\end{align}
Since $c^2-1=s^2>0$ for any $r>0$, and $q + 1 +2m>0$, the sign is determined by $q+1-2m$. Thus, entanglement requires $m<(q+1)/2$. The result in Table \ref{threshold_formulas_table} follows. 

To derive the condition for non-classicality, a similarity transformation applied to $\textbf{V}-\textbf{I}/2$ gives the matrix
\begin{align}
\left ( \begin{array}{cccc}
     A_X - 1/2&  \frac{1}{2}\sqrt{q}sa & 0 & 0\\
    \frac{1}{2}\sqrt{q}sa &(c -1)/2 & 0 & 0\\
    0 & 0 & A_P - 1/2& -\frac{1}{2}\sqrt{q}sa^{-1} \\
     0 & 0 & -\frac{1}{2}\sqrt{q}sa^{-1} & (c -1)/2
\end{array}\right),
\label{cov_PSA_reorder}
\end{align}
which has the same eigenvalues as $\textbf{V}-\textbf{I}/2$. The block structure of this matrix implies that the eigenvalues of $\textbf{V}-\textbf{I}/2$ are just the eigenvalues of the two non-zero blocks. Either block has a negative eigenvalue if and only if its determinant is negative. Of the two blocks, the upper block has a negative determinant deeper at a larger value of $n_\text{ch}$ than the lower, and therefore determines the threshold condition to be
\begin{align}
    \left(A_X-\frac{1}{2}\right)\left(\frac{c-1}{2}\right)-\frac{1}{4}qs^2a^2<0.
\end{align} This gives the result reported in Table \ref{threshold_formulas_table}.

To derive the corresponding threshold for phase-sensitive pre-amplification, apply the transformation of Eq. \ref{PSA_transform} to the measured signal quadrature, followed by the conventional amplification stage and the thermal-loss channel. The receiver quadrature is
\begin{align}
X_s^{\rm out}
=
\sqrt{\eta G_A}\,aX_s
+
\sqrt{\eta}\,X_{\rm noise}
+
\sqrt{1-\eta}\,X_{\rm th}.
\end{align}

Referring this result to the input gives
\begin{align}
\label{PSA_quad}
\widetilde X_s
=
\frac{X_s^{\rm out}}
{a\sqrt{\eta G_A}}
=
X_s+Z_{\rm PSA},
\end{align}
where
\begin{align}
\label{Z_PSA}
Z_{\rm PSA}
=
\frac{1}{a\sqrt{G_A}}X_{\rm noise}
+
\sqrt{\frac{1-\eta}{\eta G_Aa^2}}X_{\rm th}.
\end{align}
The added-noise modes are statistically independent of the original TMSV signal and idler. Hence
\begin{align}
\operatorname{MMSE}(\widetilde X_s|X_i)
=
\operatorname{MMSE}(X_s|X_i)
+
\operatorname{Var}(Z_{\rm PSA}).
\end{align}
Sub-vacuum conditional prediction requires
\begin{align}
\operatorname{MMSE}(\widetilde X_s|X_i)<\frac12 .
\end{align}
Solving this inequality for the channel thermal occupation gives the phase-sensitive threshold reported in Table \ref{threshold_formulas_table}.

\subsection{Derivation of the classical state MMSE bound}
\label{classical_state_MMSE_bound}
An arbitrary two-mode Glauber–Sudarshan \(P\)-classical state can be written
\begin{align}
\rho_{si}=
\int d^2\alpha_s\,d^2\alpha_i\,
\mathcal{P} (\alpha_s,\alpha_i)
|\alpha_s\rangle\langle\alpha_s|
\otimes
|\alpha_i\rangle\langle\alpha_i|,
\qquad \mathcal{P}\ge0 .
\end{align}
For a coherent-state component \(|\alpha_s\rangle\), the signal homodyne outcome can be represented as
\begin{align}
X_s=\mu_s+\xi_s,
\qquad
\mu_s=\sqrt{2}\operatorname{Re}\alpha_s,
\qquad
\operatorname{Var}(\xi_s)=\frac12,
\end{align}
where \(\xi_s\) represents the vacuum fluctuation associated with the coherent-state quadrature measurement. Conditioning on an idler outcome \(X_i=x_i\) changes the posterior distribution of \(\mu_s\), but does not alter this coherent-state vacuum variance. Therefore,
\begin{align}
\operatorname{Var}(X_s|X_i=x_i)
=
\frac12+
\operatorname{Var}(\mu_s|X_i=x_i)
\ge\frac12 .
\end{align}
Averaging over \(x_i\),
\begin{align}
\operatorname{MMSE}(X_s|X_i)
=
\mathbb E_{X_i}\!\left[\operatorname{Var}(X_s|X_i)\right]
\ge\frac12 .
\end{align}

\EOD
\end{document}